\documentstyle[aps,prb,epsf]{revtex} 

\begin{document}
\draft

\title
{
\LARGE \bf
Metastability of $(d+n)$-dimensional elastic manifolds}
\author{{\bf D. A. Gorokhov  and  G. Blatter}
\\{\it Theoretische Physik, ETH-H\"onggerberg,
CH-8093 Z\"urich, Switzerland}\\
e-mail: gorokhov@itp.phys.ethz.ch }
\maketitle
\begin{abstract}
We investigate the depinning of a massive elastic manifold
with $d$ internal dimensions, embedded in a $(d+n)$-dimensional space,
and subject to an isotropic pinning potential 
$V({\bf u})=V(|{\bf u}|).$
The tunneling process is driven by 
a small external
force ${\bf F}.$ We find the zero temperature and high temperature
instantons and show that for the case $1\le d\le 6$ the problem exhibits
a sharp transition from quantum to classical behavior:
At low temperatures $T<T_{c}$ the Euclidean action is constant
up to exponentially small corrections, while for $T> T_{c},$
${S_{\rm Eucl}(d,T)}/{\hbar}={U(d)}/{T}.$ The results are universal
and do not depend on the detailed shape of the trapping
potential $V({\bf u}).$
Possible applications of the problem to the depinning of vortices in
high-$T_{c}$ superconductors and nucleation in $d$-dimensional 
phase transitions
are discussed. 
In addition, we
determine the 
high-temperature asymptotics of the
preexponential factor for the $(1+1)$-dimensional problem.
\end{abstract}
\vskip0.5cm

{\hskip1.6cm preprint ETH-TH/98-10; accepted for publication in Phys. Rev. B}

\section{Introduction}
The decay of metastable states is a basic phenomenon of great
generality\cite{Haenggi} with numerous applications in a large number of contexts,
ranging from the decay of false vacua in field theory\cite{Coleman},
e.g. in cosmology\cite{Linde},
to the creep type motion of topological defects in 
solids\cite{Blatter,Petukhov}.
At a given temperature $T$ the inverse lifetime $\Gamma$
of a metastable system can be written in the form
$\Gamma =Ae^{-{S_{\rm Eucl}(T)}/{\hbar}},$ with $S_{\rm Eucl}(T)$
the Euclidean action of the saddle-point configuration and $A$ 
the prefactor determined by the associated fluctuations. 
In this paper we present a calculation of the  full temperature
dependence of the Euclidean action of a metastable 
$d$-dimensional
elastic manifold 
pinned by an isotropic potential $V({\bf u})=V(|{\bf u}|)$
in the presence of a small external force. Typical 
examples of physical realizations of this model system are
a string pinned by a linear or columnar defect 
((1+1)-, and (1+2)-dimensional problems respectively)
or a membrane
pinned by a plane ((2+1)-dimensional problem).   

With increasing temperature, the decay changes its nature from
quantum to classical.
The function $S_{\rm Eucl}(T)$ might be either 
a smooth function of temperature
or exhibit a sharp kink with a discontinuity in its first derivative at the 
crossover temperature $T_{c}$.
The former case is called a 
``second-order'' transition from quantum to classical 
behavior\cite{LO}, while in 
the latter case a first-order transition takes place.
The wording ``transition'' is appropriate as the
crossovers have all the features of mean-field phase transitions upon
identifying the Euclidean action with the free energy.

The possibility of a first-order transition in a tunneling
problem has first been discussed by Lifshitz and Kagan\cite{Lifshitz}
in the context of quantum phase transitions in ${}^{4}{\rm He}$
systems.
The general theory and an appropriate criterion for a first-order transition
in 1D Hamiltonian systems has been developed by Meshkov\cite{Meshkov}
and by Ioselevich and Rashba\cite{Ioselevich},
see also the paper of Chudnovsky\cite{Chudnovsky}. 
For the case of non-Hamiltonian systems
this problem has been considered in Ref.\onlinecite{Gorokhov}.
Recently, the possibility of observing a first-order transition
from quantum to classical behavior in spin systems has been discussed
in Ref.\onlinecite{Ch1}.

Systems with many degrees of freedom have been studied by
Garriga\cite{Garriga} (see also Ref.\cite{Ferrera}) 
and he finds that,
for the case
of tunneling of massive $d=2$ and $d=3$ elastic manifolds  
between two 
minima disbalanced by the action of 
a small external force,
the Euclidean action exhibits a discontinuous derivative,
while for a $d=1$ dimensional string the problem exhibits a smooth crossover\cite{Ivlev}.
A natural counterpart of the bistability problem
studied by Coleman, Garriga, and others is the generic problem
of weak metastability, see Fig.~1. 
In this paper we then study the problem where
an elastic manifold is pinned by an isotropic potential well and
is driven by a weak  external force.
After the tunneling process (or thermal activation at high
temperatures) the manifold becomes free, while in the previous case
it is repinned again in the second valley. 
The approximation introduced by Coleman\cite{Coleman}
and  used by Garriga is 
usually termed the  ``thin wall approximation'', while in the  present problem it is more appropriate to denote the technique used as a
``thick wall approximation''.

We find that at least for the case $1\le d\le 6$ the problem 
exhibits a sharp transition from quantum to classical behavior:
The function $S_{\rm Eucl}(d,T)$ is a constant (up to
exponentially small corrections) at temperatures lower than
the critical temperature $T_{c},$ while for $T>T_{c},$
${S_{\rm Eucl}(d,T)}/{\hbar}={U(d)}/{T},$ with $U$ the activation energy.
The (1+1)-dimensional problem 
has been considered by Skvortsov\cite{Skvortsov}.
 In the present paper
we generalize the results of this work for an arbitrary number
of internal degrees of freedom $d$ and improve on the calculation of
the preexponential factor for the $(1+1)$-problem as
the approximations made in Ref.\onlinecite{Skvortsov} turn out to be too rough
to reach the correct result.

\centerline{\epsfxsize=9cm \epsfbox{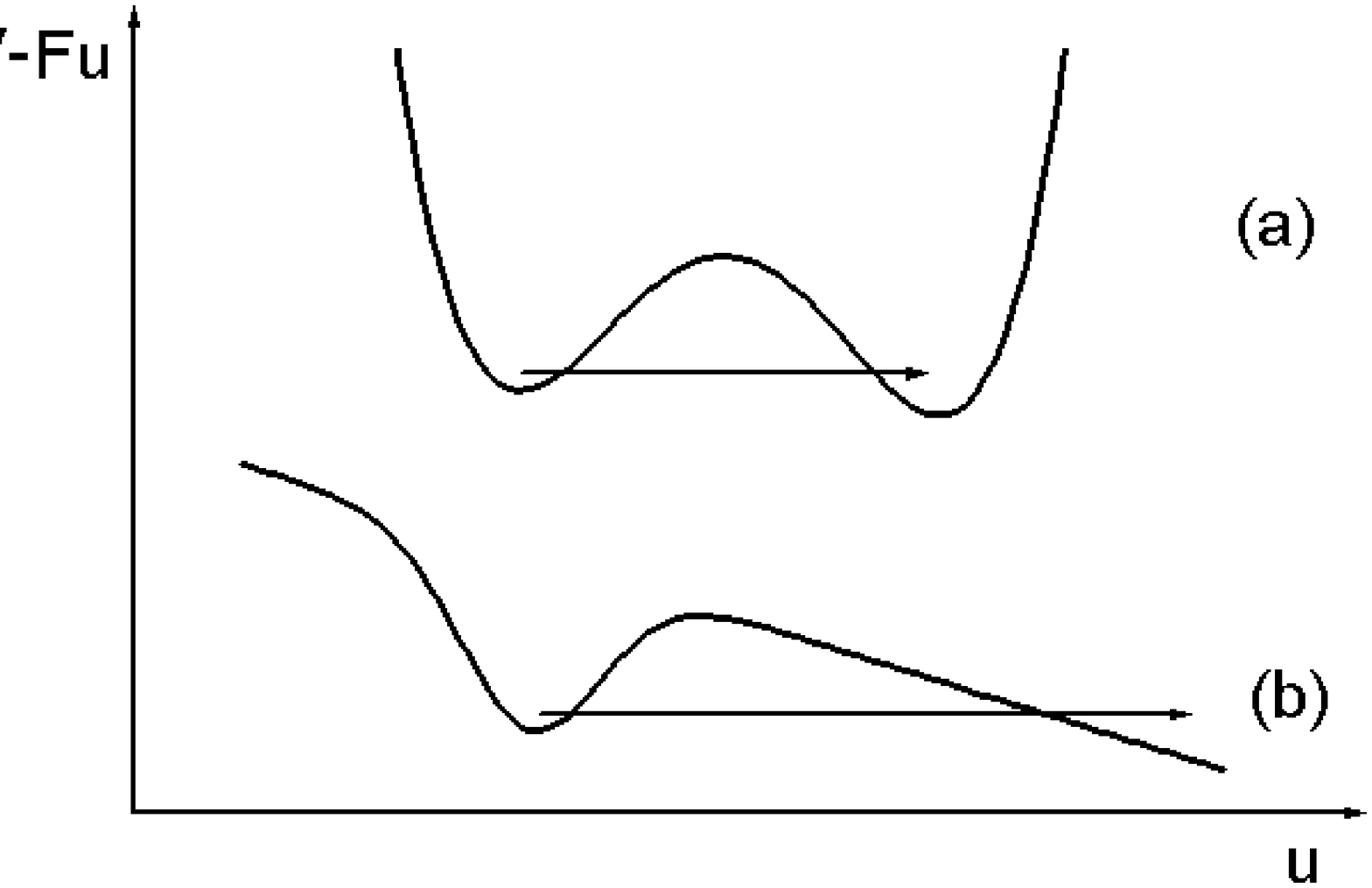}}
\vskip1cm
{ {\bf Fig.1}~The two generic decay problems of
bistability (a) and weak metastability (b).}
\vskip0.5cm
\hskip-0.7cm

 The paper is organized as follows: in Sec.~\ref{Ea}
we formulate the problem. In Sec.~\ref{quantuminst} and \ref{thth}
the zero temperature and thermal instantons are found and the 
corresponding Euclidean
actions are calculated.
In Sec.~\ref{instability} the instability temperature
of the thermal instanton is calculated and the full temperature
dependence of the Euclidean action is found. 
Sec.~\ref{preexponential} is devoted to the calculation of the 
high-temperature asymptotics of the preexponential factor
for the $(1+1)$-problem.
We summarize the results 
in Sec.~\ref{conclusion}
and discuss their possible applications
to the thermal depinning of vortices in high-$T_{c}$ superconductors
and nucleation phenomena in phase transitions.

\section{Euclidean action at low and high temperatures}
\label{Euclid}
\subsection{General expression}
\label{Ea}
Consider a massive manifold with $d$ internal dimensions
embedded in a $(d+n)$-dimensional space. The (real time)
Lagrangian of the manifold can be written in the form
\begin{equation}
{\cal L}[{\bf u}({\bf x},t)]=
\int\limits_{\Omega} 
\thinspace d{\bf x}
\left \{ \frac{\rho}{2}
{\left (\frac{\partial {\bf u}}{\partial t}\right )}^{2}-
\frac{\epsilon}{2}
{\left (\frac{\partial {\bf u}}{\partial {\bf x}}\right )}^{2}
-V(|{\bf u}|)+{\bf F}\cdot {\bf u}\right \}.
\label{Lagrangian}
\end{equation}
Here, ${\bf u}\in {\cal R}^{n}$ is the (transverse) displacement field
of the manifold, ${\bf r}\in {\cal R}^{d}$ is the vector 
characterizing the internal degrees of freedom,
$\rho$ and $\epsilon$ are
 the mass density and the elasticity of the manifold, respectively, and
$V(|{\bf u}|)$ is an $O(n)$-invariant trapping potential.
The function $V(u)$ is supposed to be monotonously 
increasing, $V(0)=0, V(\infty )=V_{0}.$ 
The integration in Eq.~(\ref{Lagrangian}) goes over
the volume $\Omega$ of the manifold.
Finally, ${\bf F}=(F,0,\dots ,0)$ is the external driving force
which we assume to be small, $F\ll {V_{0}}/{a},$
with $a$ the characteristic radius of the pinning potential $V(u).$
Specific realizations of this model system are 
strings moving
 in a plane ($(1+1)$-problem) or in space ($(1+2)$-problem),
or  membranes pinned by planar
boundaries or interfaces  in 3D space  ($(2+1)$-problems).

After the transformation 
$t\rightarrow -i\tau$ and $S\rightarrow -iS_{\rm Eucl}$ 
the Euclidean action of
the manifold reads
\begin{equation}
S_{\rm Eucl}[{\bf u}({\bf x},\tau)]=
\int\limits_{-{\hbar}/{2T}}^{+{\hbar}/{2T}}
d\tau\int\limits_{\Omega}
\thinspace d{\bf x}
\left \{ \frac{\rho}{2}
{\left (\frac{\partial {\bf u}}{\partial \tau}\right )}^{2}+
\frac{\epsilon}{2}
{\left (\frac{\partial {\bf u}}{\partial {\bf x}}\right )}^{2}
+V(|{\bf u}|)-{\bf F}\cdot {\bf u}\right \}.
\label{Euclidinitial}
\end{equation}
After introducing the new variables 
${\bf x}^{\prime}={{\bf x}}/{\sqrt{\epsilon}}$ and 
${\tau}^{\prime}={\tau}/{\sqrt{\rho}}$,
the Euclidean action (\ref{Euclidinitial}) can be 
rewritten in the form
\begin{equation}
S_{\rm Eucl}[{\bf u}({\bf x}^{\prime},{\tau}^{\prime})]=
{\rho}^{{1}/{2}}{\epsilon}^{{d}/{2}}
\int\limits_{-{\hbar}/{2\sqrt{\rho}T}}^{+{\hbar}/{2\sqrt{\rho}T}}
{d\tau}^{\prime}\int\limits_{{\Omega}^{\prime}}\thinspace d{\bf x}^{\prime}
\left \{ \frac{1}{2}
{\left (\frac{\partial {\bf u}}{\partial {\tau}^{\prime}}\right )}^{2}+
\frac{1}{2}
{\left (\frac{\partial {\bf u}}{\partial {\bf x}^{\prime}}\right )}^{2}
+V(|{\bf u}|)-{\bf F}\cdot {\bf u}\right \}.
\label{Euclideanfinal}
\end{equation}
Below we shall work with this Euclidean action and find its zero-
and finite temperature instantons. 

\subsection{Zero temperature instanton}
\label{quantuminst}
Let us calculate the Euclidean action corresponding to the 
extremal trajectories at a given temperature $T.$
The variation 
of the action (\ref{Euclideanfinal}) yields
the saddle-point equation 
\begin{equation}
\frac{\partial^{2}{\bf u}}{\partial {{\tau}^{\prime}}^{2}}
+\Delta^{\prime}{\bf u}=
\frac{\partial V}{\partial {\bf u}} -{\bf F}.
\label{saddlepoint}
\end{equation} 
We need to find the solution satisfying the boundary condition
${\bf u}({\bf x}^{\prime},0)=
{\bf u}({\bf x}^{\prime},{\hbar}/{\sqrt{\rho}}T).$ 
The solution of Eq.~(\ref{saddlepoint})
can be written in the form ${\bf u}=\left (u,\dots ,0\right ).$
At zero temperature the instanton is spherically symmetric
and introducing the new variable 
$r^{2}={{t}^{\prime}}^{2}+{{\bf r}^{\prime}}^{2}$
we obtain an ordinary differential equation
for the function $u(r),$
\begin{equation}
\frac{1}{r^{d}}{\left (r^{d}{u}^{\prime}\right )}^{\prime}=
u^{\prime\prime}+\frac{d}{r}u^{\prime}=
\frac{\partial V}{\partial u}-F.
\label{saddlenew}
\end{equation} 
The function $u(r)$ must be 
non-zero and
continuous, have a continuous derivative
$u^{\prime}(r),$ and satisfy the boundary condition
\begin{equation}
u(r),u^{\prime}(r)\rightarrow 0, \ r\rightarrow\infty.
\label{boundarycondition}
\end{equation}
The corresponding solution consists of two 
parts: a ``macroscopic'' solution for $r\alt R_{d}$
($R_{d}$ is a parameter which will be determined later)
and a ``microscopic'' solution for $r\agt R_{d}.$
The former corresponds to displacements $u\agt a$
(we remind the reader that $a$ is the characteristic radius of the 
pinning potential), while the latter describes the bounce 
in the region $u\alt a,$ 
where the potential is relevant. 

Consider the     
``microscopic'' solution where we can neglect
the (small) force $F$ in the equation of motion~(\ref{saddlenew})
as compared to the potential term ${\partial_{u}}V$.
Furthermore,
in the limit of small $F$ the radius $R_{d}$ of the nucleus is large
(we will check this assumption at the end), 
such that we can neglect the term ${d u^{\prime}}/{r}$
in Eq.~(\ref{saddlenew}), and the remaining equation takes the form
\begin{equation}
u^{\prime\prime}=\frac{\partial V}{\partial u}.
\label{Newton}
\end{equation}
Eq.~(\ref{Newton}) is identical to  
Newton's second law. Accounting for the boundary condition
(\ref{boundarycondition}) we obtain
\begin{equation}
\frac{{u^{\prime}}^{2}}{2}-V(u)=0,
\label{conservationlaw}
\end{equation} 
 which is equivalent
to the law of energy conservation for the 1D conservative motion
in classical mechanics.
As $V(u\agt a)\simeq V_{0}$,
we can obtain the boundary condition for the function
$u(r)$
\begin{equation}
u^{\prime}(R_{d})=-\sqrt{2V_{0}}.
\label{boundary}
\end{equation}
A simple integration then provides the microscopic part
of the solution, once the parameter
$R_{d}$ is known.

Next, we solve Eq.~(\ref{saddlenew}) in the region
$r<R_{d}$ with the boundary conditions given by Eq.~(\ref{boundary})
and $u(R_{d})\sim a.$ As we will see below, $u(0)\propto {1}/{F},$
allowing us to use the condition $u(R_{d})=0$
without changing the 
main term in the asymptotics of the Euclidean action
in the limit $F\rightarrow 0.$ 
 In the region $r<R_{d}$ 
we can neglect the term ${\partial V}/{\partial u}$
in Eq.~(\ref{saddlenew}) and
the general solution can be written in the form
\begin{equation}
u(r)=-\frac{Fr^{2}}{2(d+1)}+C_{1}+C_{2}f(r), 
\end{equation}
where
\begin{equation}
f(r)=
\left\{ \begin{array}{r@{\quad\quad}l} 
\displaystyle{\frac{1}{r^{d-1}}}, 
& {d\ne 1,}\\ \noalign{\vskip 5 pt}
\displaystyle{\ln{ r}}, & 
{d=1.}\end{array}\right.
\label{fx}
\end{equation}
The term $C_{2}f(r)$ is always singular at the point
$r=0$,
hence $C_{2}=0.$ Taking into account the boundary condition
(\ref{boundary}) we obtain
\begin{equation}
R_{d}=\frac{\sqrt{2V_{0}}(d+1)}{F}
\label{Rd}
\end{equation} 
and the quantum instanton can be written in the form
\begin{equation}
u(r)=\frac{F}{2(d+1)}\left (R_{d}^{2}-r^{2}\right ).
\label{instanton}
\end{equation}
Substituting the function $u(r)$ into Eq.~(\ref{Euclideanfinal})
and taking into account that 
$r^{2}={\tau^{\prime}}^{2}+{{\bf x}^{\prime}}^{2}$
we obtain the zero temperature 
Euclidean action 
\begin{equation}
S_{\rm Eucl}(T=0,d)=
\frac{2^{{(d+3)}/{2}}A_{d+1}}{\left (d+3\right )}
{\left (d+1\right )}^{d}
{\left (\frac{\sqrt{\epsilon V_{0}}}{F}\right )}^{d}
\sqrt{\rho V_{0}}\frac{V_{0}}{F}
+O\left (\frac{1}{F^{d}}\right ),
\label{Sq}
\end{equation}
where $A_{d}={2\pi^{{d}/{2}}}/{\Gamma{({d}/{2})}}$ is the 
surface area of a unit sphere in $d$-dimensional space.

Neglecting the terms
$F$ or ${\partial V}/{\partial u}$ in Eq.~(\ref{saddlenew})
might seem questionable
as there is an interval with the length of order $u_{F}$ where 
${\partial V}/{\partial u}\bigg |_{u_{F}}\simeq F.$
However, if this interval is much smaller than $R_{d}$
the approximation used above is valid: 
Assume that
the potential $V(u)$ at large $u$ behaves as
$V(u)\simeq V_{0}-{B}/{u^{\alpha}},\thinspace\alpha >0.$
The equation $V^{\prime}=F,\thinspace F\rightarrow 0$
can be easily solved and yields the result 
$u_{F}\simeq {\left ({\alpha B}/{F}\right )}^{{1}/{(\alpha +1)}},$
i.e., in the limit $F\rightarrow 0,$ $u_{F}\ll R_{d}$
and the approximation made above is asymptotically correct.

\subsection{Thermal instanton}
\label{thth}
Next, let us calculate the activation energy. The saddle-point
solution is time-independent in this case, i.e., we
can eliminate the term 
${\partial^{2}{\bf u}}/{\partial {{\tau}^{\prime}}^{2}}$ in 
Eq.~(\ref{saddlepoint}). The remaining equation is identical to the
$(d-1)$-dimensional zero temperature problem considered above and
we can use the instanton (\ref{instanton})
with $d\rightarrow d-1.$ 
We substitute this 
solution into Eq.~(\ref{Euclideanfinal}) and obtain the expression for
the activation energy,
\begin{equation}
U(d)=
\frac{2^{{(d+2)}/{2}}A_{d}}{\left (d+2\right )}
{d}^{d-1}
{\left (\frac{\sqrt{\epsilon V_{0}}}{F}\right )}^{d}V_{0}
+O\left (\frac{1}{F^{d-1}}\right ).
\label{U}
\end{equation}
We remind the reader that for the thermal instanton 
$S_{\rm Eucl}(T\gg T_{c},d)={\hbar U(d)}/{T}.$

Next, let us calculate the two characteristic temperatures
$T_{c}(d)$ and $T^{*}(d).$ The former corresponds to the value of $T,$
where the quantum and thermal exponents ${S_{\rm Eucl}(d)}/{\hbar}$
and ${U(d)}/{T}$ match up, while the latter
defines the limit of applicability of the zero temperature instanton
(see Fig.~2):
at temperatures $T>T^{*}(d)$ the instanton~(\ref{instanton})
does not satisfy the periodic
boundary 
condition ${\bf u}({\bf x},0)={\bf u}({\bf x}, {\hbar}/{T}).$
Note that the thermal instanton is always a valid solution as it is
time-independent and satisfies the periodic boundary condition
identically.  
 
The exponents ${S_{\rm Eucl}(d)}/{\hbar}$ and ${U(d)}/{T}$ are equal at the temperature
$T_{c}(d)$ given by the expression
(we make use of Eqs.~(\ref{Sq}) and (\ref{U}))
\begin{equation}
T_{c}(d)=
\frac{1}{\sqrt{2\pi}}\frac{d+3}{d(d+2)}
{\left (\frac{d}{d+1}\right )}^{d}
\frac{\Gamma (\frac{d+1}{2})}{\Gamma (\frac{d}{2})}
\frac{\hbar F}{\sqrt{\rho V_{0}}}.
\label{Tc}
\end{equation}

Formally, the quantum instanton found in Section~\ref{quantuminst}
is valid only at zero temperature, however, we can use
it also at $T\ne 0$ 
as long as the periodic boundary conditions are satisfied
asymptotically. 
The zero temperature bounce solution then can be still applied
if the periodicity in imaginary time $\tau^{\prime},$ 
${\hbar}/{\sqrt{\rho}\hskip0.1cm T}$,
is larger than the diameter of the nucleus $2R_{d},$
i.e., if   
$R(d)<{\hbar}/{2\sqrt{\rho}T}$
(see Fig.~2).

\centerline{\epsfxsize=7cm \epsfbox{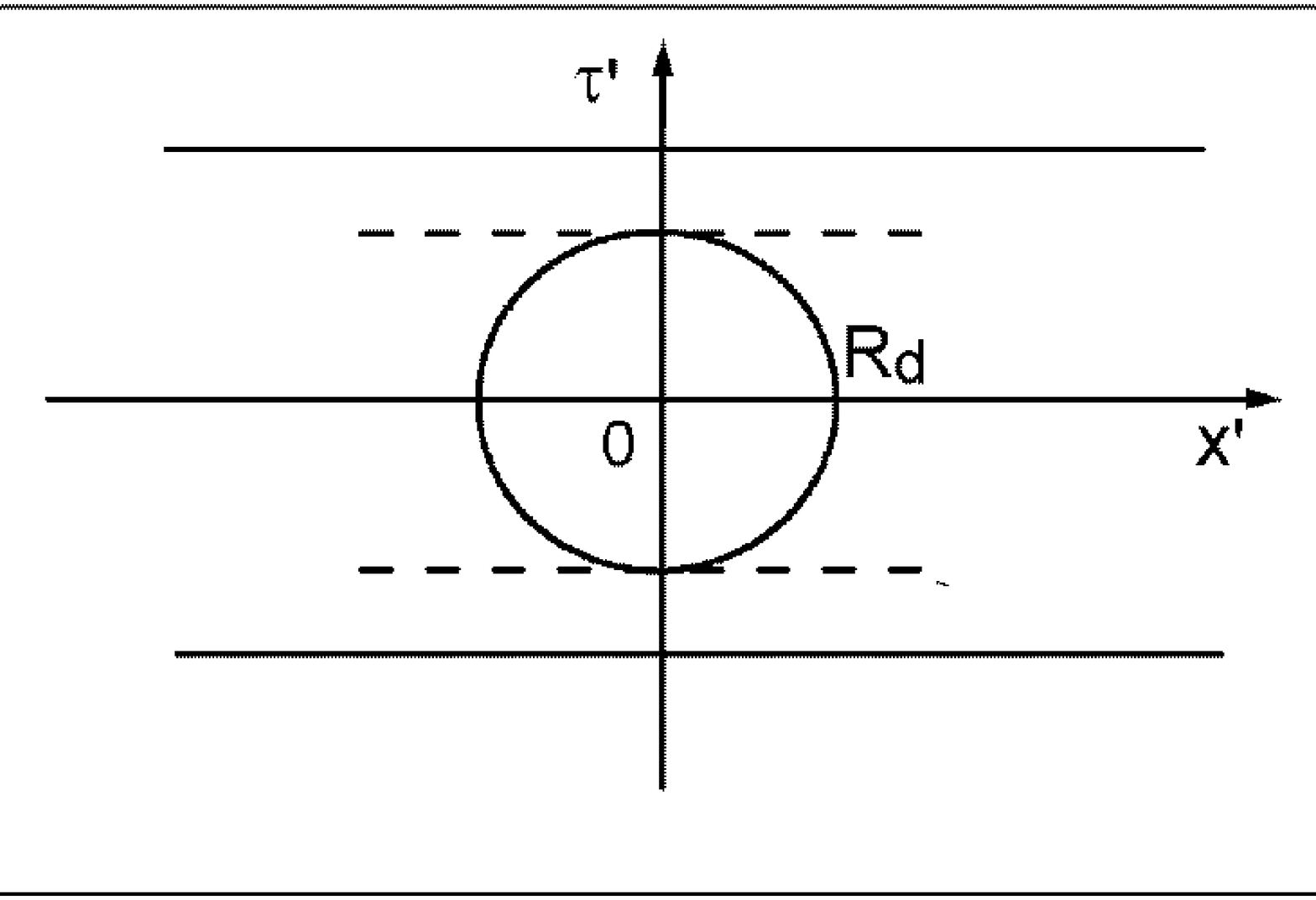}}
\vskip-3.2cm
{
\footnotesize
 {\bf Fig.2}~Nucleus corresponding to the low temperature decay process.
As long as  the diameter $2R_{d}$  remains
 smaller than the periodicity in time 
${\hbar}/{\sqrt{\rho}T}$ 
(solid lines)
 the instanton (\ref{instanton})
asymptotically satisfies
the boundary conditions.
At temperatures $T>T^{*}(d)={\hbar}/{2\sqrt{\rho}R_{d}}$
the boundary conditions are not fulfilled 
and the distortion of the nucleus' walls becomes relevant.}
\vskip0.5cm
Making use of Eq.~(\ref{Rd}) we obtain the following
expression for the temperature $T^{*}(d)$ at which one needs
to take into account the distortion of the quantum instanton
due to the periodic boundary conditions,
\begin{equation}
T^{*}(d)=
\frac{1}{2\sqrt{2}(d+1)}
\frac{\hbar F}{\sqrt{\rho V_{0}}}.
\label{T*}
\end{equation}

\section{Instability Temperature and Euclidean action}
\label{instability}
In this section we calculate the instability temperature $T_{0}(d)$ of 
the thermal instanton. At this temperature
the saddle-point solution becomes time-dependent. Then we show
that $T_{0}(d)<T_{c}(d)<T^{*}(d)$ for $1\le d\le 6$
and, consequently, the problem exhibits a first-order transition
from quantum to classical behavior.

Let us expand the Euclidean action around the thermal saddle-point
solution. As the bounce satisfies the Euler-Lagrange equations,
the term proportional to $\delta{\bf u}$ ($\delta{\bf u}$
describes the distortion of the saddle-point configuration)
is equal to zero. To second order in $\delta{\bf u}$ we  obtain
\begin{eqnarray}
S_{\rm Eucl}[{\bf u}_{th}({\bf x}^{\prime})+\delta{\bf u}] & \simeq &
S_{\rm Eucl}[{\bf u}_{th}({\bf x}^{\prime})]+
\frac{1}{2}
\int\limits_{-{\hbar}/{2\sqrt{\rho}T}}^{+{\hbar}/{2\sqrt{\rho}T}}\thinspace
d{\tau}^{\prime}
\int\limits_{\Omega^{\prime}}\thinspace d{\bf x^{\prime}}
\delta u_{1}
\left ({\hat H}_{1}-\rho
\frac{\partial^{2}}{\partial{\tau^{\prime}}^{2}}\right )
\delta u_{1}
\nonumber\\
& + &
\frac{1}{2}
\int\limits_{-{\hbar}/{2\sqrt{\rho}T}}^{+{\hbar}/{2\sqrt{\rho}T}}\thinspace
d{\tau}^{\prime}
\int\limits_{\Omega^{\prime}}\thinspace d{\bf x^{\prime}}
\delta {{\bf u}_{\bot}}
\left (
{ {\hat H}_{\bot}}-\rho\frac{\partial^{2}}{\partial{\tau^{\prime}}^{2}}
\right )
\delta {{\bf u}_{\bot}},
\end{eqnarray}
where $\delta u_{1}=\delta{\bf u}\cdot (1,0\dots 0)$
and $\delta {{\bf u}_{\bot}}=\delta {\bf u}-\delta {u}_{1}\cdot (1\dots 0).$
The operator ${\hat H}_{1}$ can be written in the 
form
\begin{equation}
{\hat H}_{1}=
-\Delta^{\prime} +\frac{\partial^{2}V}{\partial u^{2}}\bigg |_{u_{th}}.  
\label{eigenvalue}
\end{equation}
The instability temperature is determined by its
only negative eigenvalue. The operator ${\hat H}_{\bot}$
has only positive eigenvalues.

Let us calculate the lowest eigenvalue 
$-\omega^{2}$
of the operator~(\ref{eigenvalue}).
Here, we follow the procedure described in detail in 
Refs.\onlinecite{Garriga,Ivlev,Kramer}.
 First, we show that the
operator~(\ref{eigenvalue}) has $d$ zero eigenvalues:
the $d$ functions $\phi_{i}={\partial u_{th}}/{\partial x_{i}^{\prime}},$
$i=1,\dots ,d,$ satisfy the equation 
${\hat H}_{1}\phi_{i}=0.$
Due to the spherical symmetry of the 
instanton and, consequently, of the ``potential'' in
Eq.~(\ref{eigenvalue}), the eigenfunctions of the $d$-dimensional
Schr\"odinger operator~(\ref{eigenvalue}) can be represented 
in the form
$\phi_{i}=\Psi_{i} ({\tilde r} ) Y_{i}(\Omega )={\partial_{{\tilde r}} u_{th}}
{\partial_{x_{i}^{\prime}}{\tilde r}},$
with $\Psi_{i} ({\tilde r} )$ and $Y_{i}(\Omega )$
the radial and angular components, respectively, 
${\tilde r}^{2}=\sum_{i=1}^{d}{x_{i}^{\prime}}^{2}.$
For the $d$ eigenfunctions corresponding to the zero eigenvalue
$\Psi_{i} ({\tilde r} )=
\Psi({\tilde r} )={\partial_{{\tilde r}} u_{\rm th}}
\propto {\tilde r}$ for ${\tilde r}<R_{d-1}$
(see Eq.~(\ref{instanton}) and the first paragraph
in section~\ref{thth}).
In the region ${\tilde r}\agt R_{d-1}$ the eigenfunctions
of the lowest levels are asymptotically ($F\rightarrow 0$)
equal,
allowing us to find the boundary condition for the lowest
eigenvalue $-\omega^{2}$ from that of the zero eigenvalue wavefunction.
The condition for binding  
the radial parts of the
wavefunctions
at the point ${\tilde r} =R_{d-1}$ takes the form
\begin{equation} 
\frac{\Psi^{\prime} ({\tilde r} )}{\Psi}
\Bigg |_{{\tilde r} =R_{d-1}}=\frac{1}{R_{d-1}}.
\label{binding}
\end{equation}
The general solution of the equation
\begin{equation}
{\hat H}_{1}\Psi_{-1}=-\omega^{2}\Psi_{-1}
\end{equation}
 in the region
${\tilde r}<R_{d}$ can be written in the form
\begin{equation}
\Psi_{-1}({\tilde r})=
{\tilde r}^{{(2-d)}/{2}}
\left [
C_{1}I_{{(d-2)}/{2}}(\sqrt{\omega}{\tilde r} )+C_{2}K_{{(d-2)}/{2}}
(\sqrt{\omega}{\tilde r} )
\right ],
\end{equation}
where $I_{\nu}(z)$ and $K_{\nu}(z)$
are Bessel functions of imaginary argument. 
The function $K_{{(d-2)}/{2}}(z)$ is singular at the point $z=0$
and hence $C_{2}=0.$
(For the case $d=1$ this analysis is not valid, however
it can be easily shown that the equation for the lowest eigenvalue
(see Eq.~(\ref{mu}))
remains the same). 
 Using Eq.~(\ref{binding})
we obtain a transcendental equation for
the negative eigenvalue 
$-\omega^{2}(d)=-{4\pi^{2}{\rho} T_{0}^{2}(d)}/{{\hbar}^{2}}$,
\begin{equation}
\frac{\mu_{d}I^{\prime}_{{(d-2)}/{2}}(\mu_{d})}
{I_{{(d-2)}/{2}}(\mu_{d})}=\frac{d}{2}, \ \ \ \mu_{d}=\omega (d)R_{d-1},
\label{mu}
\end{equation}
and
\begin{equation}
T_{0}(d)=\frac{\mu_{d}}{2\sqrt{2}\pi d}
\frac{\hbar F}{\sqrt{\rho V_{0}}}.
\label{temp0}
\end{equation}
Using Eqs.~(\ref{Tc}) and (\ref{T*}) and the definition of the parameter
$\mu_{d},$ see Eq.~(\ref{mu}), we find
\begin{equation}
\chi_{d}\equiv\frac{T_{c}}{T^{*}}=
\frac{2}{\sqrt{\pi}}
\frac{(d+1)(d+3)}{(d+2)d}
{\left (\frac{d}{d+1}\right )}^{d}
\frac{\Gamma (\frac{d+1}{2})}{\Gamma (\frac{d}{2})}
\end{equation}
and 
\begin{equation}
\frac{T_{0}(d)}{T_{c}(d)}=
\frac{d+1}{d}\frac{\mu_{d}}{\pi \chi_{d}}.
\end{equation}
\begin{table}
\caption{Ratios ${T_{c}(d)}/{T^{*}(d)}$ and ${T_{0}(d)}/{T_{c}(d)}$
for $1\le d\le 6.$ For ${T_{0}(d)}<{T_{c}(d)}<{T^{*}(d)}$
the system exhibits a sharp transition from quantum to classical behavior.
For $d=7,$ both ratios exceed unity and no conclusion
can be drawn regarding 
the order of the transition.}
\vskip0.5cm
\begin{tabular}{l l l l}
dimensionality, d & ${\hskip0.4cm \mu_{d}}$ 
& {\hskip-0.4cm $\chi_{d}={T_{c}(d)}/{T^{*}}(d)$} 
& {\hskip-0.4cm ${T_{0}(d)}/{T_{c}(d)}$}\\ 
\noalign{\vskip 3 pt}\hline\noalign{\vskip 3 pt}
{\hskip1.3cm 1}& 1.1997& $\frac{8}{3\pi}\approx 0.8488$ & 0.8998 \\ 
\noalign{\vskip 3 pt}\hline\noalign{\vskip 3 pt}
{\hskip1.3cm 2}& 1.6083& $\frac{5}{6}\approx 0.8333$ & 0.9215 \\ 
\noalign{\vskip 3 pt}\hline\noalign{\vskip 3 pt}
{\hskip1.3cm 3}& 1.9150& $\frac{27}{10\pi}\approx 0.8594$ & 0.9457 \\ 
\noalign{\vskip 3 pt}\hline\noalign{\vskip 3 pt}
{\hskip1.3cm 4}& 2.1725& $\frac{112}{125}\approx 0.8960$ & 0.9647 \\ 
\noalign{\vskip 3 pt}\hline\noalign{\vskip 3 pt}
{\hskip1.3cm 5}& 2.3993& $\frac{5000}{1701\pi}\approx 0.9357$ & 0.9795 \\ 
\noalign{\vskip 3 pt}\hline\noalign{\vskip 3 pt}
{\hskip1.3cm 6}& 2.6048& $\frac{32805}{33614}\approx 0.9759$ & 0.9918 \\ 
\end{tabular}
\vskip0.5cm
\label{temperatures}
\end{table}

In Table~\ref{temperatures} we 
summarize the numerical values of
the parameters
${\chi}_{d}$ and ${T_{0}(d)}/{T_{c}(d)}.$ We see
that for $1\le d\le 6$ the relation 
$T_{0}(d)< T_{c}(d)<T^{*}(d)$ holds 
and, consequently, the Euclidean action
 has the form plotted in Fig.~3: It is equal to 
$S_{\rm Eucl}(d)$ (see Eq.~(\ref{Sq})) for $T<T_{c}(d)$ (see Eq.~(\ref{Tc})),
whereas for $T>T_{c}(d)$ we have $S_{\rm Eucl}(d,T)={\hbar U(d)}/{T}$.
The jump from the zero-temperature 
to the high-temperature instanton takes place at $T=T_{c}$:
the system exhibits a sharp transition from quantum
to classical behavior, see Fig.~3. 
\centerline{\epsfxsize=9cm \epsfbox{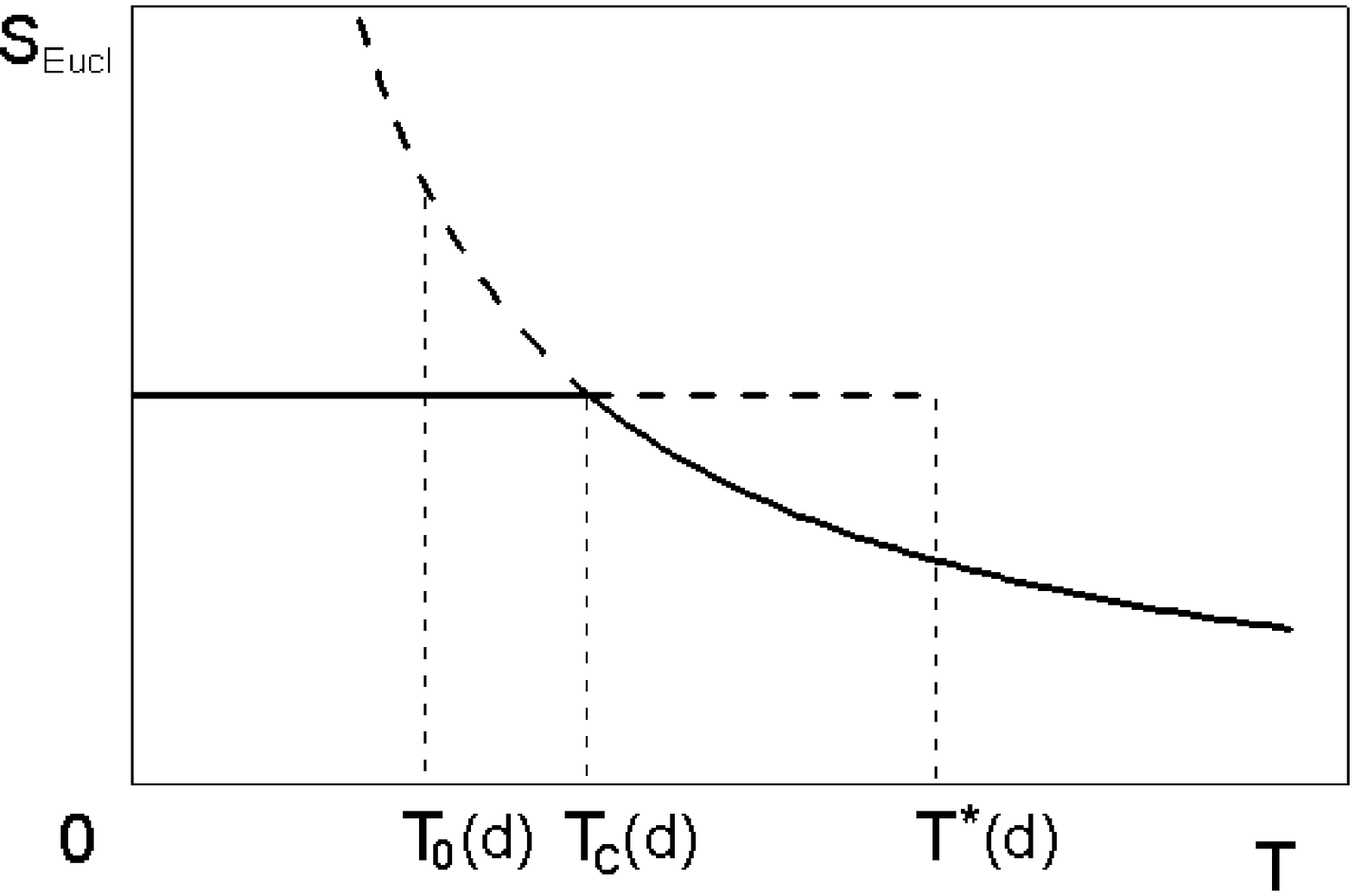}}
\vskip0.5cm
{
\footnotesize
 {\bf Fig.3}~The Euclidean action as a function of temperature
for $1\le d\le 6.$ For $T<T_{c}$ the Euclidean action is constant
and given by Eq.~(\ref{Sq}) (see Eq.~(\ref{Tc})
for an expression for $T_{c}$).
For $T>T_{c}$ the action is equal to ${\hbar U(d)}/{T}$, with
$U(d)$ given by Eq.~(\ref{U}).
}
\vskip0.5cm

\section{Preexponential factor for the (1+1)-problem}
\label{preexponential}
In this section we calculate the 
preexponential factor
for the $(1+1)$-problem in the high temperature region,
improving on the results of Ref.\onlinecite{Skvortsov}.
We start from
Langer's expression for the decay 
rate\cite{Langer1} (see also Ref.\onlinecite{Affleck})
\begin{equation}
\hbar\Gamma=2T_{0}(1)
\frac{{\rm Im}Z}{Z},
\label{Langer}
\end{equation}
where $T_{0}(1)=T_{0}(d=1),$ see Eq.~(\ref{temp0})
and Table~\ref{temperatures},
and $Z$ is the partition function of the system.
For a (1+1)-string we can write\cite{Ivlev,Rajaraman}
\begin{equation}
\hbar\Gamma=
\frac{T_{0}(1)L}{\sqrt{2\pi T}}
{\left [\int\limits_{-\infty}^{+\infty}
dx{\left (\frac{\partial u_{th}}{\partial x}\right )}^{2}
\right ]}^{{1}/{2}}
{\left |
\frac{{\rm det}\left ({\delta^{2}S}/{\delta u^{2}}|_{u=0}\right )}
{{\rm det}^{\prime}\left ({\delta^{2}S}/{\delta u^{2}}|_{u=u_{th}(r)}\right )}
\right |}^{{1}/{2}}\exp{\left (-\frac{U(1)}{T}\right )},
\label{DECAY}  
\end{equation} 
where 
$L$ is the length of the string and
the prime indicates that we exclude the zero
eigenvalue of the operator 
${\hat H}_{1}={\delta^{2}S}/{\delta u^{2}}|_{u=u_{th}(x)}$.
The activation energy $U(1)$ is given by the expression
\begin{equation}
U(1)=\frac{4\sqrt{2}}{3}
\sqrt{\epsilon V_{0}}\frac{V_{0}}{F}
\label{acten}
\end{equation}
and the eigenvalues of the operators
${\hat H}_{0}\equiv{\delta^{2}S}/{\delta u^{2}}|_{u=0}=
-\partial_{x}^{2}+\partial_{u}^{2}V |_{u=0}$ and
 ${\hat H}_{1}\equiv
{\delta^{2}S}/{\delta u^{2}}|_{u=u_{th}(x)}=
-\partial_{x}^{2}+
\partial_{u}^{2}V |_{u_{th}}$
take the form
\begin{equation}
\lambda_{0,\alpha n}=
\lambda_{0,\alpha}+\rho{\left (\frac{2\pi T}{\hbar}n\right )}^{2},\ 
n=0,\pm 1,\pm 2\dots
\label{eig0}
\end{equation}
and
\begin{equation}
\lambda_{1,\alpha n}=
\lambda_{1,\alpha}+\rho{\left (\frac{2\pi T}{\hbar}n\right )}^{2},\ 
n=0,\pm 1,\pm 2\dots ,
\label{eig1}
\end{equation}
respectively.
Carrying out the product over $n$ we obtain the high temperature result
\begin{equation}
\hbar\Gamma=
\frac{T_{0}(1)L}{\sqrt{2\pi T}}
{\left [\int\limits_{-\infty}^{+\infty}
dx{\left (\frac{\partial u_{th}}{\partial x}\right )}^{2}
\right ]}^{{1}/{2}}
{\left |
\frac{{\rm det}{\hat H}_{0}}
{{\rm det}^{\prime}{\hat H}_{1}}
\right |}^{{1}/{2}}\exp{\left (-\frac{U^{*}(1)}{T}\right )},
\label{DECAY1} 
\end{equation}
where 
\begin{equation}
U^{*}(1)=
U(1)-\frac{\hbar}{2\pi\sqrt{\rho}}
\left [\int_{0}^{k^{*}}\thinspace dk\thinspace \delta\rho (k)
\sqrt{\kappa +\epsilon k^{2}}
-\sum_{\alpha} \sqrt{\lambda_{1,\alpha}}
\right ]
\label{renorm}
\end{equation}
 is the quantum renormalized activation
energy, see Ref.\onlinecite{Ivlev}.
In Eq.~(\ref{renorm}) $\kappa =V^{\prime\prime}(0),$
$\delta\rho (k)$ is the difference in the 
continuous part of the 
spectral densities
of the operators ${\hat H}_{0}$ and ${\hat H}_{1}$,
and the sum is taken over the positive discrete eigenvalues of the operator
${\hat H}_{1}$ (note that the operator ${\hat H}_{0}$ has 
no discrete eigenvalues).
 At large $k^{*}$ the correction to
$U(1)$ diverges as $\ln k^{*}$, i.e.,  
we have to introduce a cutoff $k^{*}$
into the problem.
The determinant ratio in Eq.~(\ref{DECAY1}) has been calculated by 
Kr\"amer and Kuli\a'c\cite{Kramer} using the Gelfand-Yaglom
formula\cite{Gelfand,Kleinert} with 
the result
\begin{equation}
\left |
\frac{{\rm det}{\hat H}_{0}}
{{\rm det}^{\prime}{\hat H}_{1}}
\right |\sim
\frac{F}{a}\exp{\left (\frac{\sqrt{2\kappa V_{0}}}{F}\right )}.
\end{equation}
The coefficient of proportionality depends on the detailed form of the 
pinning potential $V(u).$ The final answer for the decay rate
then takes the form
\begin{equation}
\Gamma\sim FL
{\left (\frac{1}{\rho aT }\sqrt{\frac{V_{0}}{\epsilon}}\right )}^{{1}/{2}}
\exp{\left \{\frac{\sqrt{2\kappa V_{0}}}{F}-
\frac{U^{*}(1)}{T}\right \}};
\label{otwet}
\end{equation}
accounting for gaussian fluctuations we obtain an
exponential enhancement of the decay rate. 
Note that
$U^{*}\sim {1}/{F}$,
see Eqs.~(\ref{renorm}) and (\ref{acten}).
The result~(\ref{otwet}) is valid for 
temperatures $T$ below the kink activation energy $E_{k}$ 
\begin{equation}
T\alt\frac{FU^{*}(1)}{\sqrt{2\kappa V_{0}}}
\sim a\sqrt{\epsilon V_{0}}\sim E_{k}
\label{quantumfl}
\end{equation}
(here we assumed that the quantum correction to the activation energy
$U(1)$ is small and $\kappa\sim {V_{0}}/{a^{2}}$).
At higher temperatures we need to take into account
the renormalization of the free energy due to thermal fluctuations:
The free energy per unit length of an elastic string trapped in
a potential well coincides with the ground state energy 
of a massive particle in the equivalent quantum mechanical 
problem\cite{Nelson}, where the temperature plays the role of
Planck's constant. If the temperature is small,
the ground state energy is close to zero. At high temperatures, 
satisfying the condition $V_{0} \alt {T^2}/{\epsilon a^2}$
(which is the condition for a weak potential well, see 
Ref.~\onlinecite{Landau}), the ground state energy
increases substantially and becomes comparable with the potential
well depth. In this case the thermal instanton
found in Sec.~\ref{Euclid} is not applicable and the problem
is more conveniently solved in the quantum mechanical 
formulation: Since in the 1D quantum particle trapped in
an attractive potential always exhibits a bound state, the
$(1+1)$-dimensional string problem at $F=0$ does not undergo 
a depinning transition at any temperature.

It is instructive to compare the result (\ref{otwet})
to that obtained for the case of thermally activated motion
of a string between two nearly degenerate 
minima\cite{Ivlev,Marchesoni,Marchesoni1}, see Fig.~1(a),
where the prefactor shows a power-like dependence on
the external force $F$. The different dependences of the prefactors
follow from the different low energy spectra: In the present case we have
$\sim{V_{0}}/{Fa}$ low-lying eigenvalues of order ${F^{2}}/{V_{0}}$
(see Ref.\cite{Gorokhov1}),
the product $P(F)$ of which behaves as $\ln P(F)\sim -{1}/{F}$,
while for the problem studied in 
Refs. \onlinecite{Ivlev,Marchesoni,Marchesoni1}
only one small eigenvalue exists\cite{Gorokhov}.

Finally, let us compare the result~(\ref{otwet}) with that obtained
previously by Skvortsov\cite{Skvortsov}
where the semiclassical approximation for the eigenvalues
of the operators ${\hat H}_{0}$  and ${\hat H}_{1}$ has been used.
The derivation in Ref.\onlinecite{Skvortsov} differs 
from the present one in two respects: {\it i}) The calculation
 has been based on the low temperature expression
\begin{equation}
\hbar\Gamma =2T\frac{{\rm Im}Z}{Z}
\label{nultemperatur}
\end{equation} 
for the decay rate, such that the final high temperature result 
lacks the correct classical limit 
($\Gamma\propto {1}/{\hbar}$, see Eq.~(57) 
in Ref.\onlinecite{Skvortsov}).
{\it ii}) Correcting for the factor ${T_{0}}/{T}$,
the remaining difference can be 
 traced back to the use of 
the quasiclassical approximation in the calculation of the ratio
(\ref{otwet}), which produces a prefactor $\propto{F}^{{3}/{2}}$
rather than the correct result $\propto F$.
Note that at zero temperature Eq.~(\ref{nultemperatur}) is applicable,
however, the semiclassical approach is not accurate enough
to produce
the correct expression for the decay rate
(though it is still
  possible to obtain the correct exponential enhancement
of the preexponential factor and the correct quantum renormalized
Euclidean action $S^{*}$, 
see Ref.\onlinecite{Skvortsov}, similar to the high temperature case
discussed above). 
In principle, the complete prefactor 
in the low temperature quantum regime
can be caclulated  
using the procedure suggested in Refs.\onlinecite{Selivanov,Kiselev}:
The calculation of the determinant ratios can be reduced to 
the calculation of an infinite product
of determinants of 1D Schr\"odinger operators,
which then can be calculated numerically. 

\section{Conclusion}
\label{conclusion}
As possible applications of the present problem we discuss 
{\it i)} the thermal depinning of vortices in high-$T_{c}$ superconductors
 and {\it ii)} the nucleation of phase transitions in the vicinity of a 
boundary or an interface.

{\it i)} The dynamics of vortices 
in high-temperature superconductors
is dominated either by the dissipative or the Hall term.
The vortex mass can be neglected. However at high temperatures
the activation energy is independent of the dynamics (see Eq.~(\ref{U}))
and we can use Eq.~(\ref{U}) for the description 
of the thermal  depinning of a single vortex line ($d=1$) 
from a columnar defect ($n=2$),
or of a vortex sheet ($d=2$) from an interface between 
superconducting grains\cite{IK} 
($n=1$). We wish to point out that Eq.~(\ref{U})
can also be used for the activation energy of a single vortex
depinning from a twin boundary.
Note that a vortex pinned by a columnar defect ($(1+2)$-dimensional problem)
again does not undergo a depinning transition at any finite temperature:
the corresponding 2D quantum mechanical problem always has a bound 
state and hence the vortex remains localized at all temperatures provided
that $F=0$, see Ref.\onlinecite{Nelson}. As our result is
restricted to temperatures below the kink energy, our analysis
is not in contradiction with this result.

{\it ii)} Consider the nucleation
of a new phase\cite{Lifshitz} in the vicinity of a 
boundary\cite{Skvortsov}or interface. 
Let us assume that there exist 
two competing ($D$-dimensional) phases,
the stability of which are depending on the given external conditions.
Initially, phase 1 is prepared,
after which the external conditions are changed adiabatically,
rendering phase 1 metastable.
The nucleation of phase
2 takes a finite time and might be of two types:
usual bulk nucleation or nucleation in the vicinity of the boundary.
The latter can be described as the problem of depinning of a 
$\left ((D-1)+1\right )$-dimensional manifold.
The boundary between the two phases plays the role
of the elastic manifold, while the difference in the chemical potentials
plays the role of the small external force.
It is interesting to study the competition between these two
possible types of nucleation. 
The free energy ${\tilde U}_{D}$ 
of the $D$-dimensional spherical
nucleus of radius $R$ is given by the expression
\begin{equation}
{\tilde U}_{D}(R)=\epsilon A_{D}R^{D-1}-
FV_{D}R^{D},
\end{equation}
where $A_{D}={2\pi^{{D}/{2}}}/{\Gamma ({D}/{2})}$
and $V_{D}={2\pi^{{D}/{2}}}/{D\Gamma ({D}/{2})}$
are the surface and the volume of a unit sphere
in $D$-dimensional space. Minimizing this energy with respect to $R$
we obtain
\begin{equation}
R=\frac{\left (D-1\right )\epsilon }{F}
\end{equation}
and
\begin{equation}
{\tilde U}_{D}=
\frac{2\pi^{{D}/{2}}}{\Gamma ({D}/{2})}
\frac{{\left (D-1 \right )}^{D-1}}{D}
\epsilon
{\left (\frac{\epsilon}{F}\right )}^{D-1}.
\end{equation}
This expression should be compared to $U_{D}\equiv U(D-1)$
as given by  Eq.~(\ref{U}).
After some algebra we obtain
\begin{equation}
\frac{{\tilde U}_{D}}{U_{D}}=
\frac{\sqrt{\pi}}{2^{{(D+1)}/{2}}}
\frac{(D^{2}-1)}{D}
\frac{\Gamma\left ({(D-1)}/{2}\right )}
{\Gamma\left ({D}/{2}\right )}
{\left (\frac{\epsilon}{V_{0}}\right )}^{{(D+1)}/{2}}.
\end{equation}
It is reasonable to assume that $\epsilon\approx V_{0}$,
i.e., the boundary tension $\epsilon$ roughly matches up
with the pinning potential $V_{0}$.
 In this case
${{\tilde U}_{D}}/{U_{D}}>1$ for $D=2$ and 3, i.e.,
nucleation at the boundary is more favorable
than that in the bulk (the same result applies also for the weak
pinning case $V_{0}\ll \epsilon$).
The analysis for the quantum case  is more
complicated: The kinetic energy
of the nucleus can be written in the form
$f(R){\dot R}^{2},$ with $f(R)$ a function of
the nucleus' radius $R$.
The process of nucleation of the new phase leads to a redistribution
of mass in space as the phases have different densities in
general. 
For the case $D=3$
the function $f(R)$ has been calculated in Ref.\onlinecite{Lifshitz},
where it has been assumed that
both phases  
are incompressible
quantum liquids. For the case $D=2$ this approach gives
a divergent kinetic energy and, consequently, we
need a more detailed description of the phases 1 and 2
to reach a sensible result.
Similarly, the 
effective mass per area of the boundary between the two phases  
cannot be calculated  
assuming incompressible phases.

Briefly summarizing, we have analyzed the problem 
of depinning of a $(d+n)$-dimensional massive elastic manifold
from an $O(n)$-invariant trapping potential 
$V({\bf u})=V(|{\bf u}|)$ in the presence of a small
external force ${\bf F}.$ For the case $1\le d\le 6$ the Euclidean
action has been calculated in the whole temperature range,
see Eq.~(\ref{Sq})  for $T<T_{c}$
($T_{c}$ is given by Eq.~(\ref{Tc})) and Eq.~(\ref{U}) for
$T>T_{c}$,  
and we have found
a sharp transition from quantum to classical behavior, see Fig.~3. The 
high-temperature asymptotics for the preexponential factor
of the $(1+1)$-problem has been calculated.
Possible applications to the thermal depinning
of vortices in high-$T_{c}$ superconductors and the
nucleation problem
in the vicinity of a boundary
 in $D$-dimensional phase transitions have been discussed and
we have shown that for $D=2-4$  nucleation at the boundary is
more favorable than from the bulk. 

\acknowledgments
We thank Thomas Christen for helpful discussions.

\end{document}